\documentclass[a4paper, amsfonts, amssymb, amsmath, reprint, showkeys, nofootinbib, twoside]{revtex4-1}
\usepackage[english]{babel}
\usepackage[utf8]{inputenc}
\usepackage[colorinlistoftodos, color=green!40, prependcaption]{todonotes}

\usepackage{amsthm}
\usepackage{mathtools}
\usepackage{xcolor}
\usepackage{graphicx}
\usepackage[left=23mm,right=13mm,top=35mm,columnsep=15pt]{geometry} 
\usepackage{adjustbox}
\usepackage{placeins}
\usepackage[T1]{fontenc}
\usepackage{lipsum}
\usepackage{csquotes}

\definecolor{journalLinks}{rgb}{0,0,0.5}
\newcommand{\MYhref}[3][blueLinks]{\href{#2}{\color{#1}{#3}}}

\usepackage[pdftex, pdftitle={Article}, pdfauthor={Author}]{hyperref} % For hyperlinks in the PDF
\def\etal{{\it et al.\/}}

\begin{document}
%\title{Arrival time delay of astroparticles with different modified dispersion relations in Finsler spacetime}
\title{Trajectories of astroparticles in pseudo-Finsler spacetime with the most general modified dispersion}\thanks{J.~Zhu, B.-Q.~Ma, Eur. Phys. J. C 83 (2023) 349\\~\url{https://doi.org/10.1140/epjc/s10052-023-11517-8}}

\author{Jie Zhu}
    %\email[Correspondence email address: ]{email@institution.com}% Your name
    \affiliation{School of Physics,
    %and State Key Laboratory of Nuclear Physics and Technology, 
    Peking University, Beijing 100871, China}
    
\author{Bo-Qiang Ma}
    \email[Correspondence email address: ]{mabq@pku.edu.cn}% Your name
    \affiliation{School of Physics,
    %and State Key Laboratory of Nuclear Physics and Technology, 
    Peking University, Beijing 100871, China\\%}
    %\affiliation{
    Center for High Energy Physics, Peking University, Beijing 100871, China\\%}
    %\affiliation{
    Collaborative Innovation Center of Quantum Matter, Beijing, China}

%\date{\today} % Leave empty to omit a date

\begin{abstract}
Finsler geometry is a natural and fundamental generalization of Riemann geometry, and is a tool to research Lorentz invariance violation.
% The Finsler structure depends on both coordinates and velocities. 
We find the connection between the most general modified dispersion relation and a pseudo-Finsler structure, and then we calculate the arrival time delay of astroparticles with different modified dispersion relations in the framework of Finsler geometry. 
The result suggests that the time delay is irrelevant with the exact form of the modified dispersion relation. 
If the modified term becomes 0 when $E=p$, there is no arrival time difference, otherwise the time delays only depend on the Lorentz violation scale and the order at which the Lorentz invariance breaks.
%but related with the order at which the Lorentz invariance breaks and whether the leading order term is 0 when $E=p$.
\end{abstract}


\maketitle

\section{Introduction} \label{sec:introduction}

As a basic symmetry of space-time, Lorentz invariance has played important roles
in various fields of physics, and it can be ranked as the crystallization of human wisdom in understanding space-time. However, 
quantum gravity (QG) phenomenology covers a wide range of subjects~\cite{quamtum}, and one of the most important QG effects is the Lorentz violation (LV)~\cite{HeMa}. 
There are many kinds of theoretical models including LV effects:
(i) Quantum gravity theory, which aims at solving the conflict between the standard model and general relativity, such as string theory\cite{string}.
(ii) Space-time structure theory, which constructs new models from the perspective of space-time structure. These theories include the very special relativity (VSR)~\cite{VSR} and the doubly special relativity (DSR)~\cite{DSR1-1,DSR1-2,DSR2-1,DSR2-2}. 
Later research suggests that the VSR is a kind of Finsler special relativity~\cite{VSRFinsler}, and also the DSR can be incorporated into the framework of Finsler geometry~\cite{DSRFinsler}.
(iii) Effective theory with extra-terms, such as the standard-model extension (SME)~\cite{SME}. 
The connection between SME and Finsler geometry has been studied in recent years~\cite{SMEFinsler}.
A common feature of many studies of LV is the introduction of modified dispersion relations (MDRs).
Girelli {\em et al.}~\cite{Girelli} proposed a possible relation between MDRs and Finsler geometry to account for the nontrivial structure of Planckian spacetime.

The above facts imply that new physics may be connected with Finsler geometry. 
In fact, many kinds of Finsler geometry are studied to pursue new physics~\cite{Foster, Silva, Pfeifer, Hohmann, Silva2}.
Also, the connection between Finsler geometry and gravitation has been studied~\cite{Chang, PfeiferW}.
Finsler and Finsler-like cosmology is also studied~\cite{Saridakis2013,Saridakis2019-1,Saridakis2019-2,Saridakis2021}, aimed to solve the problem of dark energy, dark matter, inflation, and the bounce cosmology. Ideally, physicists hope to derive the full theory from basic assumptions in a top-down way, such as those in Refs.~\cite{Saridakis2013,Saridakis2019-1,Saridakis2019-2,Saridakis2021}. 
As the circumstance of applying Finsler geometry to LV studies, we hope that we can construct a well-defined LV theory from basic assumptions on Finsler structure, thus we can derive the value of the Lorentz-violation scale $E_{\rm{LV}}$.
However, till now we know little about how we can constrain the Finsler structure to fit it with LV studies.
Thus we adopt the bottom-up method, which begins with MDRs and then researches on the corresponding Finsler structure and the follow-up consequences.

An advantage of Finsler geometry is that we can discuss the trajectories of particles.
In Finsler spacetime, the observed trajectories are identified with the geodesics of the Finsler geometry.
A large source of information about the physical properties of spacetime is obtained by observing the motion of point particles.
In fact, many works tested LV form high-energy photons~\cite{Ellis, shaolijing, zhangshu, xu1, xu2, jie} and ultrahigh-energy neutrinos~\cite{Jacob:2006gn, Amelino-Camelia:2015nqa, Amelino-Camelia:2016fuh, Amelino-Camelia:2016ohi, Huang1, Li,Huang:2022nsv}, by the arrival time differences between high-energy and low-energy particles from the same source. 
For the MDR as the form
\begin{equation}\label{eq:MDR}
    E^2=m^2+p^2+\alpha p^{n+2},
\end{equation}
where $n$ we call the broken order here, and $\alpha$ is a parameter with mass dimension and $[\alpha]=-n$,
Jacob and Piran suggested a time difference formula between a high-energy astroparticle and a normally low-energy photon in the standard model of cosmology as~\cite{Jacob}
\begin{equation}
    \Delta t=\alpha E_\mathrm{obs}^{n}\frac{1+n}{2 H_{0}} \int_{0}^{z} \frac{\left(1+z^{\prime}\right)^{n} \mathrm{~d} z^{\prime}}{\sqrt{\Omega_{m}\left(1+z^{\prime}\right)^{3}+\Omega_{\Lambda}}},\label{eq:Jacob}
\end{equation}
where $z$ is the redshift of the source of the two particles, $E_\mathrm{obs}$ is the observed energy of the high-energy particle from Earth equipment, $\Omega_{\rm{m}}$ and $\Omega_{\rm{\Lambda}}$ are universe constants, and $H_0$ is the current Hubble parameter.
In our previous research~\cite{JieFinsler}, we reconsidered the time difference problem in Finsler spacetime, the result suggested that for the MDR as Eq.~(\ref{eq:MDR}), the arrival time difference formula is surprisingly the same as Eq.~(\ref{eq:Jacob}).
However, in different models, the dispersion relations are not necessarily the same form as Eq.~(\ref{eq:MDR}).
For example, the dispersion relation in DSR-1~\cite{DSR1-1,DSR1-2} is 
\begin{equation}
    E^2=m^2+p^2+\alpha Ep^2,
\end{equation}
while in DSR-2~\cite{DSR2-1,DSR2-2} the dispersion relation is
\begin{equation}\label{eq:MDRDSR2}
    \frac{E^2-p^2}{(1-\lambda E)^2}=m^2,
\end{equation}
and series Eq.~(\ref{eq:MDRDSR2}) in the leading order of $\lambda$ the dispersion relation becomes
\begin{equation}\label{eq:MDRDSR2exp}
    E^2=m^2+p^2-2\lambda E(E^2-p^2).
\end{equation}
Different kinds of MDRs may bring different time delay formulas. 
For example, previous researches such as~\cite{Jafari} suggest that the MDR of DSR-2 as Eq.~(\ref{eq:MDRDSR2}) brings no time delays.
In the framework of Finsler geometry, different MDRs mean that the corresponding geometries are different, and thus the geodesic equations and trajectories of particles are different. 
So in Finsler spacetime, the time delays corresponding to different MDRs should be considered carefully.

\section{A Brief Introduction to Finsler Geometry}

Finsler geometry~\cite{textbook} is a natural and fundamental generalization of Riemann geometry.
For a manifold $M$, denoted by $T_x M$ the tangent space at $x \in M$, and by $T M$ the tangent bundle of $M$. Each element of $T M$ has the form $(x, y)$, where $x \in M$ and $y \in T_x M$. The natural projection $\pi: T M \rightarrow M$ is given by $\pi(x, y) \equiv x$.
A Finsler structure, or Finsler norm of $M$ is a function
\begin{equation}
    F: T M \rightarrow[0, \infty)
\end{equation}
with the following properties:
(i) Regularity: $\mathrm{F}$ is $C^{\infty}$ on the entire slit tangent bundle $T M \backslash 0$.
(ii) Positive homogeneity : $F(x, \lambda y)=\lambda F(x, y)$ for all $\lambda>0$.
(iii) Strong convexity: The $n \times n$ Hessian matrix
\begin{equation}
    g_{i j} \equiv\left(\frac{1}{2} F^2\right)_{y^i y^j}
\end{equation}
is positive-definite at every point of $T M \backslash 0$, where we have used the notation ()$_{y^i}=$ $\frac{\partial}{\partial y^i}()$.
Finsler geometry has its genesis in integrals of the form
\begin{equation}
    \int_s^r F\left(x^1, \cdots, x^n ; \frac{d x^1}{d \tau}, \cdots, \frac{d x^n}{d \tau}\right) d \tau,
\end{equation}
and its geometric meaning is the distance between two points in the Finsler manifold through a certain path.
Given a manifold $M$ and a Finsler structure $F$ on $T M$, the pair $(M, F)$ is called as a Finsler manifold. It is obvious that the Finsler structure $F$ is a function of $\left(x^i, y^i\right)$. In the case of $g_{ij}$ depending on $x^i$ only, the Finsler manifold reduces to Riemannian manifold.

To describe the ``1 + 3'' spacetime, instead of Finsler geometry we turn to pseudo-Finsler geometry.
A pseudo-Finsler metric is said to be locally Minkowskian if at every point there is a local coordinate system, such that $F = F (y)$ is independent of the position $x$.
For a massive particle propagating in 1+3 spacetime, its action can be expressed as
\begin{equation}
    S=m\int F\left(x^\mu,y^\mu\right)d\tau,
\end{equation}
where $m$ is the mass of the particle, and $y^\mu=\frac{dx^\mu}{d\tau}$ is the 4-speed of the particle.
And thus the Lagrangian of the particle is
\begin{equation}
    L=m F\left(x^\mu,y^\mu\right).
\end{equation}

In this work we focus on the geodesic equation of Finsler geometry. The geodesic equation for the Finsler manifold is given as~\cite{textbook}
\begin{equation}
    \frac{d^{2} x^{\mu}}{d \tau^{2}}+2 G^{\mu}=0, \label{eq:geodesic}
\end{equation}
where 
\begin{equation}
    G^{\mu}=\frac{1}{4} g^{\mu \nu}\left(\frac{\partial^{2} F^{2}}{\partial x^{\lambda} \partial y^{\nu}} y^{\lambda}-\frac{\partial F^{2}}{\partial x^{\nu}}\right) \label{eq:spray}
\end{equation}
is called the geodesic spray coefficient. Obviously if $F$ is a Riemann metric, then 
\begin{equation}
    G^{\mu}=\frac{1}{2} \gamma_{\nu \lambda}^{\mu} y^{\nu} y^{\lambda},\label{eq:riemann}
\end{equation}
where $\gamma_{\nu \lambda}^{\mu}$ is the Riemann Christoffel symbol. We can also see that if $F$ is locally Minkowskian, then $G^\mu=0$, and the geodesic equation~(\ref{eq:geodesic}) is actually $\frac{d^{2} x^{\mu}}{d \tau^{2}}=0$.

\section{Pseudo-Finsler Structure of Particles Subject to General Forms of Lorentz Violation} \label{sec:structure}

To simplify the discussion, here we introduce the concept of homogeneous function~\cite{hf}. A function $f(\vec{x})$ is a homogeneous function when $f(\vec{x})$ satisfies 
\begin{equation}
    f(\lambda \vec{x})=\lambda^nf(\vec{x}),\label{homogeneous}
\end{equation}
and $n$ is called the degree of homogeneity, or simply the degree. 
A slightly more general form of homogeneity is called positive homogeneity, by requiring only that the above identities hold for $\lambda>0$, and allowing any real number $n$ as a degree of homogeneity. In the following we only consider the case of positive homogeneity.
Euler's homogeneous function theorem asserts that the positively homogeneous functions of degree $n$ are exactly the solution of a specific partial differential equation
\begin{equation}
    n f(\vec{x})=\sum_{i=1}^k \frac{\partial}{\partial x_i}f(\vec{x}).
\end{equation}
We can easily see that a locally Minkowskian Finsler norm $F(y)$ is a 1-order homogeneous function.

Here we consider the most general form of modified dispersion relations of broken order $n$.
The modified dispersion relations can be expressed in the most general form as the leading term of Taylor series in natural units as
\begin{equation}
p_0^2=m^2+|\vec{p}|^2+\alpha h(p_0,\vec{p}),\label{eq:general_form}
\end{equation}
where $p_0=E$ is the energy of the particle, and $\alpha$ is a parameter with mass dimension and $[\alpha]=-n$. We can easily see that $h(p_0,\vec{p})$ is a homogeneous function of order $n+2$, which means that $h(\lambda p_0,\lambda\vec{p})=\lambda^{n+2}h(p_0,\vec{p}).$

Girelli {\em et al.}~\cite{Girelli} provided a workflow to construct Finsler norms corresponding to dispersion relations.
However, the algorithm in Ref.~\cite{Girelli} is complicated when dealing with a general MDR, so here we derive the pseudo-Finsler norm of the above MDR in a different way by means of the property of the homogeneous function.
According to the result of our previous work~\cite{JieFinsler}, we assume the pseudo-Finsler norm correspond to Eq.~(\ref{eq:general_form}) can be expressed as
\begin{equation}
F=\sqrt{\Delta+\alpha m^n g(y^0,\vec{y})}, 
\end{equation}
where $\Delta=\eta_{\mu \nu}y^\mu y^{\nu}=(y^0)^2-|\vec{y}|^2$, $y^\mu=\dot{x}^\mu=\frac{dx^\mu}{d\tau}$, and $\eta_{\mu \nu}=\text{diag}(1,-1,-1,-1)$.
The Lagrangian of the particle is
\begin{equation}
L=m F(x,\dot{x})=m\sqrt{\Delta+\alpha m^n g(y^0,\vec{y})}, 
\end{equation}
by $p_\mu=\partial L/\partial y^\mu$, in the first order of $\alpha$, we have
\begin{equation}\label{eq:p0pa}
\begin{split}
    p_0=& \frac{m y^0}{\sqrt{\Delta}}+\frac{1}{2}\alpha m^{n+1}\left(-\frac{y^0g(y^0,\vec{y})}{\Delta^\frac{3}{2}}+\frac{g_0(y^0,\vec{y})}{\sqrt{\Delta}}\right),\\
    p_a=& -\frac{m y^a}{\sqrt{\Delta}}+\frac{1}{2}\alpha m^{n+1}\left(\frac{y^ag(y^0,\vec{y})}{\Delta^\frac{3}{2}}+\frac{g_a(y^0,\vec{y})}{\sqrt{\Delta}}\right),
\end{split}
\end{equation}
where $a=1,2,3$ is a spatial index, and $g_i(y^0,\vec{y})$ means $\frac{\partial}{\partial y^i}g(y^0,\vec{y}).$
Coming Eqs.~(\ref{eq:p0pa}) and Eq.~(\ref{eq:general_form}), we get the equation for $g(y^0,\vec{y})$ in the first order of $\alpha$ as
\begin{equation}
    g(y^0,\vec{y})-y^\mu g_\mu(y^0,\vec{y})+\Delta\cdot h\left(\frac{y^0}{\sqrt{\Delta}},-\frac{\vec{y}}{\sqrt{\Delta}}\right)=0.\label{eq:eqh}
\end{equation}
Euler's homogeneous function theorem tells us that the general solution of $g(y^0,\vec{y})-y^\mu g_\mu(y^0,\vec{y})=0$ is any 1-order homogeneous function. Notice that $\Delta\cdot h(\frac{y^0}{\sqrt{\Delta}},-\frac{\vec{y}}{\sqrt{\Delta}})$ is a 2-order homogeneous function of $y^\mu$, we can easily check that $g(y^0,\vec{y})=\Delta\cdot h(\frac{y^0}{\sqrt{\Delta}},-\frac{\vec{y}}{\sqrt{\Delta}})$ is a particular solution of Eq.~(\ref{eq:eqh}). So the general solution to Eq.~(\ref{eq:eqh}) is
\begin{equation}
    g(y^0,\vec{y})=\Delta\cdot h\left(\frac{y^0}{\sqrt{\Delta}},-\frac{\vec{y}}{\sqrt{\Delta}}\right)+C(y^\mu),\label{eq:solgatF}
\end{equation}
where $C(y^\mu)$ is any 1-order homogeneous function.
However, the property of Finsler norm requests that $g(y^0,\vec{y})$ should be a 2-order homogeneous function, which means $C(y^\mu)=0$. So the pseudo-Finsler norm of the modified dispersion relations above in the first order of $\alpha$ is 
\begin{equation}
\begin{split}\label{eq:FinslerNormGeneral}
F&=\sqrt{\Delta\left[1+\alpha m^n h\left(\frac{y^0}{\sqrt{\Delta}},-\frac{\vec{y}}{\sqrt{\Delta}}\right)\right]}\\
&=\sqrt{\Delta}+\frac{\alpha m^n}{2\Delta^\frac{n+1}{2}}h\left(y^0,-\vec{y}\right),
\end{split}
\end{equation}
where $\Delta=\eta_{\mu \nu}y^\mu y^{\nu}=(y^0)^2-|\vec{y}|^2$ and $\eta_{\mu \nu}=\text{diag}(1,-1,-1,-1)$.
The result corresponds to the result obtained by Lobo and Pfeifer~\cite{Lobo}, and when $h(p_0,\vec{p})=|\vec{p}|^{n+2}$, Eq.~(\ref{eq:FinslerNormGeneral}) becomes Eq.~(21) of our previous work~\cite{JieFinsler}.

As we can see from Eq.~(\ref{eq:FinslerNormGeneral}), the pseudo-Finsler norms of particles with LV strongly rely on the forms of dispersion relations. Even if two forms of dispersion relations are of the same broken order, the corresponding pseudo-Finsler norms can be different. Thus the trajectories of particles with different MDRs are different. So it is important to discuss how different forms of MDRs influence the time delays in Finsler geometry, and this is what we discuss in the next section.

\section{Time Delay in Finsler Expanding Universe} \label{sec:time}

Here we derive the time delay of astroparticles with dispersion relations as Eq.~(\ref{eq:general_form}) in pseudo-Finsler spacetime. 
We use the same method as our previous work~\cite{JieFinsler}: First, we construct the pseudo-Finsler norm of the particle in the expanding universe; Second, we obtain the geodesic equations of the particle; Last, we solve the geodesic equations and calculate the time delay. Since the pseudo-Finsler norm is the most general and more complicated than our previous work~\cite{JieFinsler}, the derivation is more complicated.

Since $h(p_0,\vec{p})$ is a $(n+2)$-order homogeneous function of $p_\mu$, we can rewrite $h(p_0,\vec{p})$ as
\begin{equation}
    h(p_0,\vec{p})=p_0^{n+2}H\left(\frac{\vec{p}}{p_0}\right),
\end{equation}
and correspondingly,
\begin{equation}
    h(y^0,-\vec{y})=(y^0)^{n+2}H\left(-\frac{\vec{y}}{y^0}\right).
\end{equation}
Consider a particle propagating in ``1+1'' flat spacetime, the pseudo-Finsler norm can be expressed as
\begin{equation}
\begin{split}
    F=&\sqrt{(y^0)^2-(y^1)^2}\\
    &+\frac{\alpha m^n}{2}\frac{(y^0)^{n+2}}{\left(\sqrt{(y^0)^2-(y^1)^2}\right)^{n+1}}H\left(-\frac{y^1}{y^0}\right),
\end{split}
\end{equation}
where $y^0=\dot{x}^0=dt/d\tau$ and $y^1=\dot{x}^1=dx/d\tau$.
In Riemann geometry, the expanding universe can be described by the Friedmann-Robertson–Walker~(FRW) metric, and in a $1 + 1$ Riemann spacetime the length element is $ds=\sqrt{dt^2-a(t)^2dx^2}$,
%$F_R=\sqrt{(y^0)^2-(a(x^0)y^1)^2}$ in a Finsler way, 
where $a(t)$ is the cosmological expansion factor. 
In the form of Finsler structure, the corresponding pseudo-Finsler norm of FRW universe is
\begin{equation}
    F_R=\sqrt{(y^0)^2-(a(x^0)y^1)^2}.
\end{equation}
The FRW metric can be obtained by replacing the space component $d\vec{x}$ in Riemann Minkowski metric with $a(t)d\Vec{x}$,
or the space component $y^\alpha$ with $a(t)y^\alpha$ in the language of Finsler geometry. It is natural to think in this way because $a(t)$ describes how the space expands and it should be multiplied to every space component in the metric.
For the expanding pseudo-Finsler spacetime, the flat Finsler structure $ds=F(y^0,\vec{y})d\tau=F(dt,d\vec{x})$ should also be changed to
$ds=F(dt,a(t)d\vec{x})=F(y^0, a(x^0)\vec{y})d\tau$.
Thus in the expanding 1+1 spacetime the pseudo-Finsler norm can be assumed to be
\begin{equation}
\begin{split}
    F'=&\sqrt{(y^0)^2-(a(x^0)y^1)^2}\\
    &+\frac{\alpha m^n}{2}\frac{(y^0)^{n+2}}{\left(\sqrt{(y^0)^2-(a(x^0)y^1)^2}\right)^{n+1}}H\left(-\frac{a(x^0)y^1}{y^0}\right)\label{eq:Fexpand}.
\end{split}
\end{equation}
In fact, Eq.~(\ref{eq:Fexpand}) is the general case of the result of Lobe \etal~\cite{Lobo2017}, where they begin with the MDR in the expanding universe and reach the corresponding pseudo-Finsler norm.

Assume that a particle starts to move at $t=-T$ and $x=X$ with redshift $z_0$ and reaches us at $t=0$ and $x=0$, and we can measure its energy and momentum $E_\mathrm{obs}$ and $P_\mathrm{obs}$. Obviously, we have $y^0=dt/d\tau>0$, $y^1=dx/d\tau<0$, and $dx/dt<0$.
After tedious calculation (see Appendix~\ref{appendix}), the motion of the particle can be expressed as
\begin{widetext}
\begin{equation} 
\begin{split}
    \frac{dx}{dt}=&-\frac{E_\mathrm{obs}}{a\sqrt{m^2a^2+E_\mathrm{obs}^2}}+\alpha\left[
    -\frac{C_2 m^{n+2}E_\mathrm{obs}a}{C_1(m^2a^2+E_\mathrm{obs}^2)^\frac{3}{2}}
    +\frac{C_3 m^n E_\mathrm{obs}^3 a}{C_1^2 (m^2a^2+E_\mathrm{obs}^2 )^\frac{3}{2}}\right.\\
    &+\frac{a}{2(m^2a^2+E_\mathrm{obs}^2)^\frac{3}{2}}
    \int a^{-n-3}(m^2a^2+E_\mathrm{obs}^2)^{\frac{n-2}{2}} \left( (n+1) (n+2) E_\mathrm{obs}^3 (m^2 a^2+E_\mathrm{obs}^2) H\left(\frac{E_\mathrm{obs}}{\sqrt{m^2a^2+E_\mathrm{obs}^2}}\right)\right.\\
    &\left.\left.+m^2a^2 \sqrt{m^2a^2+E_\mathrm{obs}^2} \left(-m^2a^2+(2 n+1) E_\mathrm{obs}^2\right)
   H'\left(\frac{E_\mathrm{obs}}{\sqrt{m^2a^2+E_\mathrm{obs}^2}}\right)+m^4 a^4 E_\mathrm{obs} H''\left(\frac{E_\mathrm{obs}}{\sqrt{a^2
   m^2+E_\mathrm{obs}^2}}\right)\right)da\right],\label{eq:speedPm}
\end{split}
\end{equation}
\end{widetext}
where $C_1, C_2$ and $C_3$ are integration constants generated from the process of solving differential equations. 
For astroparticles, $E_\mathrm{obs}\gg m$, the first two terms with integration constants in the square brackets are suppressed in comparison to the third term in the square brackets, so we can omit the contribution of the two terms. 
We can clearly see that the equation of motion depends on the form of $H(x)$, or the form of the MDR.
However, with limit $m\rightarrow0$, surprisingly Eq.~(\ref{eq:speedPm}) has a very simple form as
\begin{equation}\label{eq:speedP}
    \frac{dx}{dt}=-\left(\frac{1}{a}+\frac{n+1}{2}\alpha H(1)E_\text{obs}^n\frac{1}{a^{n+1}}\right).
\end{equation}
If $H(1)\neq0$, we can absorb the value of $H(1)$ in $\alpha$ by define $\alpha_1=\alpha H(1)$ and $H_1(x)=H(x)/H(1)$. In other words, if $H(1)\neq0$, we can always assume $H(1)=1$, and thus Eq.~(\ref{eq:speedP}) becomes the same as Eq.~(40) of our previous work, and thus the time delay is
\begin{equation}
    \Delta t=\alpha E_\mathrm{obs}^n \frac{n+1}{2H_0}\int_0^z \frac{(1+z')^n}{\sqrt{\Omega_{m}\left(1+z^{\prime}\right)^{3}+\Omega_{\Lambda}}}dz',
\end{equation}
and is still the same formula obtained by Jacob and Piran~\cite{Jacob} in the standard model of cosmology.
If $H(1)=0$, e.g., the dispersion relation of DSR-2 as Eq.~(\ref{eq:MDRDSR2exp}) where $H(x)=-2(1-x^2)$, Eq.~(\ref{eq:speedP}) becomes 
\begin{equation}
    \frac{dx}{dt}=-\frac{1}{a}+O(\alpha^2),
\end{equation}
and the time delay becomes 
\begin{equation}
    \Delta t=O(\alpha^2),
\end{equation}
which means there is almost no time delay between astroparticles with different energies, as mentioned in Sec.~\ref{sec:introduction}. 
The result that if $H(1)=0$ then there is no time delay is almost obvious in phenomenological analysis. 
In the limit $m\rightarrow0$, $H(1)=0$ means that the modified term of the MDR turns to 0 when $E=p$, which means that the MDR can be simplified as $E=p$, just as the dispersion of normal photons, and of course no time delay appears.

The result above suggests a surprising conclusion. For modified dispersion relations as
\begin{equation}\label{eq:MDRH}
    E^2=m^2+p^2+\alpha E^{n+2} H\left(\frac{p}{E}\right),
\end{equation}
the corresponding pseudo-Finsler structures are related with $H(x)$, but the time delays calculated by solving geodesic equations in pseudo-Finsler spacetime are irrelevant with the form of $H(x)$. If $H(1)=0$, there will be no time delay, otherwise we can assume $H(1)=1$, and the formula of time delay is the same as the time delay induced by the Lorentz violation effect between two particles with different energies in the expanding universe,  i.e., Eq.~(\ref{eq:Jacob}) obtained by Jacob and Piran~\cite{Jacob} in the standard model of cosmology.

\section{Conclusion and Discussion}\label{sec:conclusion}

This work is a promotion of our previous work~\cite{JieFinsler}.
In our previous work~\cite{JieFinsler}, we derived the pseudo-Finsler structure corresponding to the MDR as Eq.~(\ref{eq:MDR}), 
and calculated the arrival time delay between astroparticles by solving geodesic equations, 
and found an interesting result that the time delay formula is the same as Jacob and Piran~\cite{Jacob} got in a different way from the standard model of cosmology.
In this work, we find an even more surprising and interesting result.
We consider a MDR with the most general form, and using the property of homogeneous function, 
we get the corresponding pseudo-Finsler norm.
By solving geodesic equations, we get the equation of motion as Eq.~(\ref{eq:speedPm}).
The equation depends on the exact form of the MDR, but interestingly, when $E_\mathrm{obs} \gg m$, 
the dependence vanishes and the time delay formula becomes the same as Eq.~(\ref{eq:Jacob}) if the modified term of MDR is not 0 when $E=p$.
For the circumstance that the modified term of MDR is 0 when $E=p$, the time delay is 0 in Finsler spacetime, just the same as the result of phenomenological analysis.

% In this work we derive the pseudo-Finsler structure of a particle subject to Lorentz violation from a more general modified dispersion relation as Eq.~(\ref{eq:general_form}). The result as Eq.~(\ref{eq:FinslerNormGeneral}) suggests that the pseudo-Finsler norm of a particle subject to Lorentz violation rely on the exact form of the modified dispersion relation. However, after a detailed calculation of the trajectory of the particle subject to Lorentz violation in the expanding universe by the geodesic equation of the pseudo-Finsler structure and calculating the arrival time delay between particles with high energy and normally low energy, we find that the time delay is irrelevant with the form of the modified term. In pseudo-Finsler spacetime, if the modified term becomes 0 when $E=p$, such as the case of DSR-2~\cite{DSR2-1,DSR2-2}, there will be no time delay between astroparticles with different energy. Otherwise the time delay only depends the order at which the Lorentz invariance breaks ($n$ in Eq.~(\ref{eq:MDRH})) and the Lorentz violation scale, and the time delay formula is the same as Jacob and Piran~\cite{Jacob} got, in a different way from the standard model of cosmology.

The result of this work provides a new perspective on the recent tests on Lorentz violation. 
Researches on Lorentz violation from high-energy photons~\cite{Ellis, shaolijing, zhangshu, xu1, xu2, jie} and ultrahigh-energy neutrinos~\cite{Jacob:2006gn, Amelino-Camelia:2015nqa, Amelino-Camelia:2016fuh, Amelino-Camelia:2016ohi, Huang1, Li, Huang:2022nsv} suggest that high-energy photons and neutrinos may be subject to Lorentz violation with broken order 1. In previous researches we may assume the dispersion relation might be $E^2=m^2+p^2+s\frac{p^3}{E_\mathrm{LV}}$, where $s=\pm1$ and $E_\mathrm{LV}$ is the Lorentz violation scale from observation. However, the result of this work suggests that the dispersion relation can be expressed as
\begin{equation}
    E^2=m^2+p^2+s\frac{E^3}{E_\mathrm{LV}}H\left(\frac{p}{E}\right), 
\end{equation}
and the only constraint on $H(x)$ is $H(1)=1$. This perspective can provide more possibilities on different models of Lorentz violation. 

\section*{ACKNOWLEDGMENTS} \label{sec:acknowledgements}
This work is supported by National Natural Science Foundation of China (Grant No.~12075003).

\begin{widetext}

\appendix*\label{appendix}
\section{Technical Details}

From Eq.~(\ref{eq:geodesic}), the geodesic equations of the pseudo-Finsler norm as Eq.~(\ref{eq:Fexpand}) can be derived as
\begin{subequations}
\label{eq:eqFRWFinsler}
\begin{equation}
\begin{split}
    \dot{y}^0&+a(x^0)a'(x^0)(y^1)^2 + \alpha m^n \frac{a'(x^0) (y^0)^{n-2}y^1}{2 \left[(y^0)^2-a(x^0)^2(y^1)^2\right]^\frac{n+2}{2}}\cdot\\
    &\cdot\left[\left((n^2+3n+2)a(x^0)^3(y^0)^2(y^1)^3-(n+2)a(x^0)(y^0)^4y^1\right)H\left(-\frac{a(x^0)y^1}{y^0}\right)\right.\\
    &+2y^0\left((n+1)a(x^0)^4(y^1)^4-(n+2)a(x^0)^2(y^0)^2(y^1)^2+(y^0)^4\right)H'\left(-\frac{a(x^0)y^1}{y^0}\right)\\
    &\left.+a(x^0)y^1\left((y^0)^2-a(x^0)^2(y^1)^2\right)^2H''\left(-\frac{a(x^0)y^1}{y^0}\right)\right]=0,
\end{split}
\end{equation}
\begin{equation}
\begin{split}
    \dot{y}^1&+2\frac{a'(x^0)}{a(x^0)}y^0y^1 + \alpha m^n \frac{a'(x^0)(y^0)^{n-1}}{2a(x^0)^2 \left[(y^0)^2-a(x^0)^2(y^1)^2\right]^\frac{n+2}{2}}\cdot\\
    &\cdot\left[n(n+2)a(x^0)^3(y^0)^2(y^1)^3H\left(-\frac{a(x^0)y^1}{y^0}\right)\right.\\
    &+\left((2n+1)a(x^0)^4y^0(y^1)^4-2(n+1)a(x^0)^2(y^0)^3(y^1)^2+(y^0)^5\right)H'\left(-\frac{a(x^0)y^1}{y^0}\right)\\
    &\left.+a(x^0)y^1\left((y^0)^2-a(x^0)^2(y^1)^2\right)^2H''\left(-\frac{a(x^0)y^1}{y^0}\right)\right]=0.
\end{split}
\end{equation}
\end{subequations}
To solve the geodesic equations at leading order in $\alpha$, following our previous work~\cite{JieFinsler}, we assume that the solution has the form
\begin{subequations}
\label{eq:form}
\begin{equation}
    y^1=\frac{C_1}{a(x^0)^2}+\alpha m^n f(\tau),\label{eq:form1}
\end{equation}
\begin{equation}
    y^0=\sqrt{\epsilon+\frac{C_1^2}{a(x^0)^2}}+\alpha m^n g(\tau), \label{eq:form2}
\end{equation}
\end{subequations}
where $C_1<0$ because $y^1=dx/d\tau<0$. Combing Eqs.~(\ref{eq:eqFRWFinsler}) and~(\ref{eq:form}), and expanding the equations to $O(\alpha^2)$, we can get the equations for $f(\tau)$ and $g(\tau)$.
We notice that $f'(\tau)=\frac{df}{da}\frac{da}{dx^0}\frac{dx^0}{d\tau}=a'(x^0)y^0\frac{df}{da}$ and the same for $g(\tau)$. Using this, we can get the equations for $f(a)$ and $g(a)$ as
\begin{subequations}
\label{eq:fg}
\begin{equation}\label{eq:f}
\begin{split}
    f'(a)&+\frac{2}{a}f(a)+\frac{(\epsilon a^2+C_1^2)^\frac{n-2}{2}}{2\epsilon^\frac{n+2}{2}a^{n+5}}\left[n(n+2)C_1^3(\epsilon a^2+C_1^2)H\left(\frac{-C_1}{\sqrt{\epsilon a^2+C_1^2}}\right)\right.\\
    &\left.+\epsilon a^2 \sqrt{\epsilon a^2+C_1^2}(\epsilon a^2-2nC_1^2)H'\left(\frac{-C_1}{\sqrt{\epsilon a^2+C_1^2}}\right)+C_1\epsilon^2a^4H''\left(\frac{-C_1}{\sqrt{\epsilon a^2+C_1^2}}\right)\right]=0,
\end{split}
\end{equation}
\begin{equation}\label{eq:g}
\begin{split}
    g'(a)&-\frac{C_1^2}{a(\epsilon a^2+C_1^2)}g(a)+\frac{2C_1}{\sqrt{\epsilon a^2+C_1^2}}f(a)+\frac{C_1(\epsilon a^2+C_1^2)^\frac{n-3}{2}}{2\epsilon^\frac{n+2}{2} a^{n+4}}\cdot\\
    &\cdot\left[-(n+2)C_1(\epsilon a^2+C_1^2)(\epsilon a^2-n C_1^2)H\left(\frac{-C_1}{\sqrt{\epsilon a^2+C_1^2}}\right)\right.\\
    &\left.2\epsilon a^2 \sqrt{\epsilon a^2+C_1^2}(\epsilon a^2-n C_1^2)H'\left(\frac{-C_1}{\sqrt{\epsilon a^2+C_1^2}}\right)+C_1\epsilon^2 a^4 H''\left(\frac{-C_1}{\sqrt{\epsilon a^2+C_1^2}}\right)\right]=0.
\end{split}
\end{equation}
\end{subequations}
The solutions to Eq.~(\ref{eq:fg}) is
\begin{subequations}
\label{eq:solfg}
\begin{equation}\label{eq:solf}
 \begin{split}
     f(a)=&\frac{C_2}{a^2}-\frac{1}{2\epsilon^\frac{n+2}{2}a^2}\int\frac{(\epsilon a^2+C_1^2)^\frac{n-2}{2}}{a^{n+3}}\left[n(n+2)C_1^3 (\epsilon a^2+C_1^2) H\left(\frac{-C_1}{\sqrt{\epsilon a^2+C_1^2}}\right)\right.\\
     &+\left.\epsilon a^2 \sqrt{\epsilon a^2+C_1^2}(\epsilon a^2-2nC_1^2)H'\left(\frac{-C_1}{\sqrt{\epsilon a^2+C_1^2}}\right)+C_1\epsilon^2a^4H''\left(\frac{-C_1}{\sqrt{\epsilon a^2+C_1^2}}\right)\right]da,
 \end{split}
\end{equation}
\begin{equation}\label{eq:solg}
\begin{split}
    g(a)=&\frac{C_3 a}{\sqrt{\epsilon a^2+C_1^2}}+\frac{C_1C_2}{a\sqrt{\epsilon a^2+C_1^2}}+\frac{C_1}{2 \epsilon^\frac{n+2}{2}a\sqrt{\epsilon a^2+C_1^2}}\cdot\\
    &\cdot\left[\epsilon a^2\int \frac{(\epsilon a^2+C_1^2)^\frac{n-1}{2}}{a^{n+3}}\left((n+2)C_1\sqrt{\epsilon a^2+C_1^2}H\left(\frac{-C_1}{\sqrt{\epsilon a^2+C_1^2}}\right)-\epsilon a^2 H'\left(\frac{-C_1}{\sqrt{\epsilon a^2+C_1^2}}\right)\right)da\right.\\
    &-\int \frac{(\epsilon a^2+C_1^2)^\frac{n-2}{2}}{a^{n+3}}\left(n(n+2)C_1^3(\epsilon a^2+C_1^2)H\left(\frac{-C_1}{\sqrt{\epsilon a^2+C_1^2}}\right)\right.\\
    &\left.\left.+\epsilon a^2 \sqrt{\epsilon a^2+C_1^2}(\epsilon a^2-2nC_1^2)H'\left(\frac{-C_1}{\sqrt{\epsilon a^2+C_1^2}}\right)+C_1\epsilon^2a^4H''\left(\frac{-C_1}{\sqrt{\epsilon a^2+C_1^2}}\right)\right)da\right].
\end{split}
\end{equation}
\end{subequations}
From Eqs.~(\ref{eq:form}) and (\ref{eq:solfg}), the ratio $dx/dt=y^1/y^0$ can be derived in the leading order of $\alpha$ as
\begin{equation}\label{eq:speed}
\begin{split}
    \frac{dx}{dt}=&
    \frac{C_1}{a\sqrt{\epsilon a^2+C_1^2}}+\alpha m^n \left[
    \frac{\epsilon C_2 a}{(\epsilon a^2+C_1^2)^\frac{3}{2}}
    -\frac{C_1C_3 a}{(\epsilon a^2+C_1^2)^\frac{3}{2}}\right.\\
    &-\frac{a}{2\epsilon^\frac{n}{2}(\epsilon a^2+C_1^2)^\frac{3}{2}}\int \frac{(\epsilon a^2+C_1^2)^\frac{n}{2}}{a^{n+3}}\left(C_1^3(n^2+3n+2) H\left(\frac{-C_1}{\sqrt{\epsilon a^2+C_1^2}}\right)\right.\\
    &\left.\left.+\frac{\epsilon a^2 (\epsilon a^2-(2n+1)C_1^2)}{\sqrt{\epsilon a^2+C_1^2}}H'\left(\frac{-C_1}{\sqrt{\epsilon a^2+C_1^2}}\right)+\frac{C_1\epsilon^2 a^4}{\epsilon a^2+C_1^2}H''\left(\frac{-C_1}{\sqrt{\epsilon a^2+C_1^2}}\right)  \right)da\right].
\end{split}
\end{equation}
Like our previous work~\cite{JieFinsler}, we let
\begin{equation}
    \epsilon=\frac{C_1^2m^2}{P_o^2},\label{eq:ep}
\end{equation}
where $[P_o]=[m]$ and $P_o>0$. Discussion in Ref.~\cite{JieFinsler} suggests that $P_\mathrm{obs}=P_o+O(\alpha^2)$, and the dispersion relation as Eq.~(\ref{eq:general_form}) suggests that $P_\mathrm{obs}=E_\mathrm{obs}+O(\alpha)$, thus here $P_o=E_\mathrm{obs}+O(\alpha)$ can be regarded as the observed energy of the particle on earth equipment. Thus we get Eq.~(\ref{eq:speedPm}).

\end{widetext}

%\appendix*
%\input{sections/appendix1.tex}
%\section{Appendix} \label{sec:appendix}


\begin{thebibliography}{4}

\bibitem{quamtum}
G.~Amelino-Camelia,
Introduction to Quantum-Gravity Phenomenology.
{Lect. Notes Phys.} {\bf 669}, {59} (2005)



\bibitem{HeMa}
For a recent review, see, e.g., P.~He, B.-Q.~Ma, Lorentz symmetry violation of cosmic photons.
\MYhref[journalLinks]{https://doi.org/10.3390/universe8060323}{Universe {8}, 323 (2022)}.
%{Universe} {\bf 8}, 323 (2022).
%URL = {https://www.mdpi.com/2218-1997/8/6/323},

\bibitem{string}
U.~Danielsson,
Introduction to string theory.
{Rep. Prog. Phys.} {\bf 64}, 51 (2001)

\bibitem{VSR}
A.~G.~Cohen, S.~L.~Glashow,
Very Special Relativity.
{Phys. Rev. Lett.} {\bf 97}, 021601 (2006).


\bibitem{DSR1-1}
G.~Amelino-Camelia, 
Testable scenario for relativity with minimum length.
{Phys. Lett. B} {\bf510}, 255 (2001).

\bibitem{DSR1-2}
G.~Amelino-Camelia, 
Relativity in space-times with short-distance structure governed by an observer-independent (Planckian) length scale.
{Int. J. Mod. Phys. D} {\bf11}, 35 (2002).

\bibitem{DSR2-1}
J.~Magueijo and L.~Smolin, 
Lorentz Invariance with an Invariant Energy Scale.
{Phys. Rev. Lett.} {\bf88}, 190403 (2002).

\bibitem{DSR2-2}
J.~Magueijo and L.~Smolin, 
Generalized Lorentz invariance with an invariant energy scale.
{Phys. Rev. D} {\bf67}, 044017 (2003).

\bibitem{VSRFinsler}
G.~W.~Gibbons, J.~Gomis, C.~N.~Pope, 
General very special relativity is Finsler geometry.
{Phys. Rev. D} {\bf 76}, 081701 (2007).
H.~F.~Goenner, G.~Y.~Bogoslovsky, 
A Class of Anisotropic (Finsler-)Space-time Geometries.
{Gen. Rel. Grav.} {\bf 31}, 1383 (1999).
A.~P.~Kouretsis, M.~Stathakopouslos, P.~C.~Stavrinos, 
General very special relativity in Finsler cosmology.
{Phys. Rev. D} {\bf79}, 104011 (2009).

\bibitem{DSRFinsler}
F.~Girelli, S.~Liberati, L.~Sindoni, 
Planck-scale modified dispersion relations and Finsler geometry.
{Phys. Rev. D} {\bf75}, 064015 (2007).

\bibitem{SME}
D.~ Colladay, V.~A~Kostelecký,
CPT violation and the standard model.
{Phys. Rev. D} {\bf55}, 6760 (1997).
D.~ Colladay, V.~A~Kostelecký,
Lorentz-violating extension of the standard model.
{Phys. Rev. D} {\bf58}, 116002  (1998).


\bibitem{SMEFinsler}
V.~A.~Kostelecký, 
Riemann-Finsler geometry and Lorentz-violating kinematics.
{Phys. Lett. B} {\bf701}, 137 (2011).
V.~A.~Kostelecký, N.~Russell, R.~Tso,
Bipartite Riemann-Finsler geometry and Lorentz violation.
{Phys. Lett. B} {\bf716}, 470 (2012).
D.~Colladay, P.~McDonald, 
Singular Lorentz-violating Lagrangians and associated Finsler structures. 
{Phys. Rev. D} {\bf92}, 085031 (2015).
N.~Russell,
Finsler-like structures from Lorentz-breaking classical particles.
{Phys. Rev. D} {\bf91}, 045008 (2015).
M.~Schreck, 
Classical Lagrangians and Finsler structures for the nonminimal fermion sector of the Standard-Model Extension.
{Phys. Rev. D} {\bf93}, 105017 (2016).
B.~R.~Edwards and V.~A.~Kostelecký, 
Riemann-Finsler geometry and Lorentz-violating scalar fields.
{Phys. Lett. B} {\bf786}, 319 (2018).
%\bibitem{Schreck1}
M.~Schreck,
Classical Lagrangians for the nonminimal Standard-Model Extension at higher orders in Lorentz violation.
{Phys. Lett. B} {\bf 793}, 70 (2019).
%\bibitem{Schreck2}
J.A.A.S.~Reis, M.~Schreck, 
Classical Lagrangians for the nonminimal spin-nondegenerate Standard-Model Extension at higher orders in Lorentz violation.
{Phys. Rev. D} {\bf 103}, 095029 (2021).

\bibitem{Girelli}
F.~Girelli, S.~Liberati, L.~Sindoni,
Planck-scale modified dispersion relations and Finsler geometry.
{Phys. Rev. D} {\bf75}, 064015 (2007).

\bibitem{Foster}
J.~Foster,  R.~Lehnert,
Classical-physics applications for Finsler $b$ space.
{Phys. Lett. B} {\bf746}, 164-170 (2015).

\bibitem{Silva}
J.E.G.~Silva, R.V.~Maluf,  C.A.S.~Almeida,
Bipartite-Finsler symmetries.
{Phys. Lett. B} {\bf798}, 135009 (2019).

\bibitem{Pfeifer}
C.~Pfeifer,
Finsler spacetime geometry in Physics.
{Int. J. Geom. Meth. Mod. Phys.} {\bf 16}, 1941004 (2019).

\bibitem{Hohmann}
M.~Hohmann, C.~Pfeifer, N.~Voicu,
Cosmological Finsler Spacetimes.
{Universe} {\bf 6}, 65 (2020).

\bibitem{Silva2}
J.E.G.~Silva,
A field theory in Randers-Finsler spacetime.
{EPL} {\bf 133}, 21002 (2021).

\bibitem{Chang}
X.~Li, Z.~Chang, 
Towards a gravitation theory in Berwald–Finsler space.
{Chinese Phys. C} {\bf34}, 28 (2010).

\bibitem{PfeiferW}
C.~Pfeifer, M.~N.~R.~Wohlfarth,
Finsler geometric extension of Einstein gravity.
{Phys. Rev. D} {\bf 85}, 064009 (2012).

\bibitem{Saridakis2013}
S.~Basilakos, A.~P.~Kouretsis, E.~N.~Saridakis, P.~C.~Stavrinos,
Resembling dark energy and modified gravity with Finsler-Randers cosmology.
{Phys. Rev. D} {\bf 88}, 123510 (2013).

\bibitem{Saridakis2019-1}
G.~Minas, E.~N.~Saridakis, P.~C.~Stavrinos, A.~Triantafyllopoulos,
Bounce Cosmology in Generalized Modified Gravities.
{Universe }{\bf 5(3)}, 74 (2019).

\bibitem{Saridakis2019-2}
S.~Ikeda, E.~N.~Saridakis, P.~C.~Stavrinos, A.~Triantafyllopoulos,
Cosmology of Lorentz fiber-bundle induced scalar-tensor theories.
{Phys. Rev. D} {\bf 100}, 124035 (2019).

\bibitem{Saridakis2021}
S.~Konitopoulos, E.~N.~Saridakis, P.~C.~Stavrinos, A.~Triantafyllopoulos,
Dark gravitational sectors on a generalized scalar-tensor vector bundle model and cosmological applications.
{Phys. Rev. D} {\bf 104}, 064018 (2021).

% \bibitem{Longo}
% M.~J.~Longo,
% Tests of relativity from SN1987A.
% {Phys. Rev. D} {\bf36}, 3276 (1987).

% \bibitem{Stodolsky}
% L.~Stodolsky, 
% The speed of light and the speed of neutrinos.
% {Phys. Lett. B} {\bf201}, 353 (1988).  

% \bibitem{method1}
% G.~Amelino-Camelia, J.~R.~Ellis, N.~E.~Mavromatos, D.~V.~Nanopoulos,
% Distance measurement and wave dispersion in a Liouville-string approach to quantum gravity.
% {Int. J. Mod. Phys. A} {\bf12}, 607 (1997).
% %{\it Int. J. Mod. Phys. A} {\bf 12} 607-624 (1997),

% \bibitem{method2}
% G.~Amelino-Camelia, J.~R.~Ellis, N.~E.~Mavromatos, D.~V.~Nanopoulos, S.~Sarkar,
% Tests of quantum gravity from observations of $\gamma$-ray bursts.
% {Nature} {\bf393}, 763 (1998).

\bibitem{Ellis}
J.~R.~Ellis, N.~E.~Mavromatos, D.~Nanopoulos, A.~S.~Sakharov, E.~K.~G.~Sarkisyan,
Robust limits on Lorentz violation from gamma-ray bursts.
{Astropart.\ Phys.\ } {\bf25}, 402-411 (2006). [Corrigendum {\bf29}, 158-159(2008)].

\bibitem{shaolijing}
L.~Shao, Z.~Xiao, B.-Q.~Ma,
Lorentz violation from cosmological objects with very high energy photon emissions.
{Astropart.\ Phys.\ } {\bf33}, 312-315 (2010).

\bibitem{zhangshu}
S.~Zhang, B.-Q.~Ma,
Lorentz violation from gamma-ray bursts.
{Astropart.\ Phys.\ } {\bf61}, 108-112 (2015).

\bibitem{xu1}
H.~Xu,B.-Q.~Ma,
Light speed variation from gamma-ray bursts.
{Astropart. Phys.}  {\bf 82}, 72 (2016).

\bibitem{xu2}
H.~Xu,B.-Q.~Ma,
Light speed variation from gamma ray burst GRB 160509A.
{Phys. Lett. B } {\bf760}, 602 (2016).

\bibitem{jie}
J.~Zhu, B.-Q.~Ma,
Pre-burst events of gamma-ray bursts with light speed variation.
{Phys. Lett. B} {\bf820}, 136518 (2021).

\bibitem{Jacob:2006gn}
U.~Jacob, T.~Piran,
Neutrinos from gamma-ray bursts as a tool to explore quantum-gravity-induced Lorentz violation.
{Nature Phys.\ } {\bf 3}, 87 (2007).

\bibitem{Amelino-Camelia:2015nqa}
G.~Amelino-Camelia, D.~Guetta, T.~Piran,
IceCube Neutrinos and Lorentz Invariance Violation.
{Astrophys.\ J.\  }{\bf 806}, 269 (2015).

\bibitem{Amelino-Camelia:2016fuh}
G.~Amelino-Camelia, L.~Barcaroli, G.~D'Amico, N.~Loret, G.~Rosati,
IceCube and GRB neutrinos propagating in quantum spacetime.
{Phys.\ Lett.\ B }{\bf 761}, 318 (2016).
	
\bibitem{Amelino-Camelia:2016ohi}
G.~Amelino-Camelia, G.~D'Amico, G.~Rosati, N.~Loret,
In-vacuo-dispersion features for GRB neutrinos and photons.
{Nat.\ Astron.\ } {\bf 1}, 0139 (2017).

\bibitem{Huang1}
%Y.~Huang, B.-Q.~Ma,
%Lorentz violation from gamma-ray burst neutrinos.
%{Communications Physics} {\bf1}, 62 (2018).
%doi:10.1038/s42005-018-0061-0
%[arXiv:1810.01652 [hep-ph]].
Y.~Huang, B.-Q.~Ma, {{Lorentz violation from gamma-ray burst neutrinos}}.
\MYhref[journalLinks]{https://doi.org/10.1038/s42005-018-0061-0}
{{Communications Physics} {\bf1}, 62 (2018)},
{doi:10.1038/s42005-018-0061-0}.
%{Comms. Phys. {1}, 62 (2018)}.
%[\MYhref[eprintLinks]{https://arxiv.org/abs/1810.01652}{{arXiv:1810.01652}}].

\bibitem{Li}
Y.~Huang, H.~Li, B.-Q.~Ma,
Consistent Lorentz violation features from near-TeV IceCube neutrinos.
{Physical Review D} {\bf 99}, 123018 (2019).

\bibitem{Huang:2022nsv}
Y.~Huang, B.-Q.~Ma, 
{{Ultra-high energy cosmic neutrinos from gamma-ray bursts}}.
\MYhref[journalLinks]{https://doi.org/10.1016/j.fmre.2022.05.022}{Fundamental Research {2} (2022) in press},
doi:10.1016/j.fmre.2022.05.022.

\bibitem{Jacob}
U.~Jacob, T.~Piran,
Lorentz-violation-induced arrival delays of cosmological particles.
{JCAP} {\bf01}, 031 (2008).

\bibitem{JieFinsler}
J.~Zhu, B.-Q.~Ma,
Lorentz-violation-induced arrival time delay of astroparticles in Finsler spacetime.
{Physical Review D}{\bf 105}, 124069 (2022)

\bibitem{Jafari}
N.~Jafari, M.~R.~R.~Good,
Dispersion relations in finite-boost DSR.
{Phys. Lett. B} {\bf 809}, 135735 (2020).


\bibitem{textbook}
D.~Bao, S.~S.~Chern, Z.~Shen,
An Introduction to Riemann–Finsler Geometry.
Graduate Texts in Mathematics 200, Springer, New York, 2000.


\bibitem{hf}
Wikipedia contributors, 
"Homogeneous function".  %https://en.wikipedia.org/w/index.php?title=Homogeneous\_function\&oldid=1109377485 
https://en.wikipedia.org/wiki/Homogeneous\_function

\bibitem{Lobo}
I.~P.~Lobo, C.~Pfeifer,
Reaching the Planck scale with muon lifetime measurements.
{Phys. Rev. D} {\bf103}, 106025 (2021).

\bibitem{Lobo2017}
I.~P.~Lobo, N.~Loret, F.~Nettel,
Rainbows without unicorns: metric structures in theories with modified dispersion relations.
{Eur. Phys. J. C} {\bf77}, 451 (2017).


\end{thebibliography}
\end{document}